\documentclass[letterpaper,twocolumn,10pt]{esapub}

\usepackage{times}
\usepackage{natbib}
\usepackage{graphicx}

\begin{document}

\title{Asteroseismology of Sun-like Stars -- A Proposal}

\author[1,2]{T.~S.~Metcalfe}
\affil[1]{Harvard-Smithsonian Center for Astrophysics, 60 Garden Street, 
          Cambridge MA 02138, USA}
\author[2]{T.~M.~Brown}
\affil[2]{High Altitude Observatory, National Center for Atmospheric 
          Research, Boulder CO 80307, USA}
\author[3]{J.~Christensen-Dalsgaard}
\affil[3]{Department of Physics and Astronomy, University of Aarhus, 
          DK-8000 Aarhus C, Denmark}

\maketitle

\begin{abstract}

In the past decade, helioseismology has revolutionized our understanding
of the interior structure of the Sun. In the next decade, asteroseismology
will place this knowledge into context, by providing structural
information for dozens of pulsating stars across the H-R diagram.  
Solar-like oscillations have already been detected from the ground in a
few stars, and several current and planned satellite missions will soon
unleash a flood of stellar pulsation data. Deriving reliable seismological
constraints from these observations will require a significant improvement
to our current analysis methods. We are adapting a computational method,
based on a parallel genetic algorithm, to help interpret forthcoming
observations of Sun-like stars. This approach was originally developed for
white dwarfs and ultimately led to several interesting tests of
fundamental physics, including a key astrophysical nuclear reaction rate
and the theory of stellar crystallization. The impact of this method on
the analysis of pulsating white dwarfs suggests that seismological
modeling of Sun-like stars will also benefit from this approach.

\end{abstract}

\section{Observational Context}

Most of what we can learn about stars comes from observations of their
outermost surface layers. We are left to infer the properties of the
interior based on our best current understanding of the constitutive
physics. The exception to this general rule arises from observations of
pulsating stars, where seismic waves probe deep through the interior and
bring information to the surface in the form of light and radial velocity
variations. The most dramatic example is the Sun, where such observations
have led to the identification of millions of unique pulsation modes, each
sampling the solar interior in a slightly different and complementary way.
The radial profile of the sound speed inferred from these data have led to
such precise constraints on the standard solar model that the observations
and theory now agree to better than a few parts per thousand over 90
percent of the solar radius \citep{jcd02}.

If we were to move the Sun to the distance of even the nearest star, most
of the pulsation modes that we now know to be present would be rendered
undetectable. We would lose most of our spatial resolution across the disk
of the star, and only those modes of the lowest spherical degree
($l\le3$) would produce significant variations in the total integrated
light or the spectral line profiles. This would reduce the number of
detectable modes from millions to dozens, leading to a corresponding
reduction in the ability of the observations to constrain the internal
structure \citep[e.g., see][]{kje99}. Even so, such data would still allow
us to determine the global properties of the star and to probe the gross
internal composition and structure, providing valuable independent tests
of stellar evolution theory.

Recent improvements in our ability to make high-precision radial velocity
measurements from the ground have been driven largely by efforts to detect
extra-solar planets. These advances in technology have simultaneously led
to the first unambiguous detections of solar-like oscillations in other
stars \citep[see][]{bk03}. Scintillation in the Earth's atmosphere
severely limits our ability to detect the corresponding light variations
due to these pulsations. But the observational requirements are similar to
programs for detecting extra-solar planet transits, so we can expect
continued rapid progress in this area---primarily from space.

\section{Computational Asteroseismology} 

Several present (WIRE, MOST) and future (COROT, Kepler) satellite missions
will soon yield nearly uninterrupted long-term coverage of many types of
pulsating stars.  We will then face the challenge of determining the
fundamental properties of these stars from the data, by attempting to
match them with the output of our computer models. The traditional
approach to this task is to make informed guesses for each of the model
parameters, and then to adjust them iteratively until an adequate match is
found. This {\em subjective} method is particularly troublesome when
combined with a {\em local} approach to iterative improvement of the
defining parameters.

An optimization scheme based on a genetic algorithm \citep{cha95,mc03} can
avoid the problems inherent in many traditional approaches. Using only
observations and the constitutive physics of the model to restrict the
range of possible values for each parameter, genetic algorithms provide a
relatively efficient means of searching globally for the optimal model.
They were inspired by Charles Darwin's notion of biological evolution
through natural selection. The basic idea is to solve an optimization
problem by {\em evolving} the global solution, starting with an initial
set of purely random guesses. The evolution takes place within the
framework of the model, and the individual parameters serve as the genetic
building blocks. Selection pressure is imposed by some goodness-of-fit
measure between the model and the observations.

\cite{met01} developed a fully parallel and distributed hardware/software
implementation of the popular {\tt PIKAIA} genetic algorithm written by
Paul Charbonneau and Barry Knapp. He used this modeling tool in the
context of white dwarf asteroseismology, leading to a number of
interesting physical results including a precise estimate of the
astrophysically important $^{12}{\rm C}(\alpha,\gamma)^{16}{\rm O}$
nuclear reaction rate \citep{met03}, and an empirical test of the theory
of stellar crystallization \citep{mmk04}. The impact of this method on the
analysis of pulsating white dwarfs suggests that seismological modeling of
other types of stars could also benefit from this approach.

We propose to extend this powerful new analysis method to treat some of
the most important problems in asteroseismology, exploiting the full
potential of the observations. After developing an interface between the
existing parallel code and models of main-sequence stars, the initial
applications will be to the well-characterized $\delta$-Scuti stars and to
the analysis of solar-like oscillations, using existing observations and
synthetic data. We describe the plans for these investigations in greater
detail below.

\subsection{Interfacing with Main-sequence Models}

The basic idea behind a genetic algorithm is fairly simple: it is just an
iterative Monte Carlo method that samples the model space randomly, but
keeps a sort of memory of what worked well in the past. It accomplishes
this through a computational analogy with the idea of biological evolution
through natural selection. It starts just like a simple Monte Carlo, where
we generate $N$ random sets of parameters, evaluate the model for each
set, and then compare them to the observations. The genetic algorithm
treats each set of parameters as an individual in a population, and
assigns each a {\it fitness} based on how well it matches the
observations. Next, it selects from this population at random, with the
fittest individuals more likely to survive. It then encodes the parameters
into simple strings of numbers, sort of like chromosomes; it pairs them up
and performs operations that are analogous to breeding and mutation, and
then decodes the strings back into numerical values for the parameters.
This produces a new population, so we evaluate the model for each case
again, and continue the whole process until some termination criterion is
met. Although it may seem a rather contorted way of exploring the model
space, there is a firm theoretical basis known as the {\it schema theorem} to
explain why it actually works in practice \citep{gol89}.

{\bf Automating the code.} Although genetic algorithms are often more
efficient than other comparably global optimization methods, they are
still quite demanding computationally. Fortunately, the procedure is
inherently parallelizable; we need to calculate many models, and each one
of them is independent of the others. So the number of available
processors determines the number of models that can be calculated in
parallel. Also, there is very little communication overhead; parameter
values are sent to each processor, and they return either a list of
pulsation periods or just a goodness-of-fit measure if the computed
periods have already been compared to the observations. The parallel
version of the {\tt PIKAIA} genetic algorithm is perfectly general, and
will not require any structural modifications to accommodate the
application to main-sequence models.

Some customization will be required, however, for the main-sequence models
that we adopt for this project. J.~Christensen-Dalsgaard's models were
developed for the analysis of helioseismic data, and were used to produce
`Model S' of \cite{jcd96}, which has been used extensively as a reference
model for inversions. Using these models for the analysis of pulsations in
Sun-like stars will provide a certain degree of internal consistency in
our understanding of solar-like oscillations.

Many of the required modifications will be similar to what was necessary
for the white dwarf code. The main challenge will be to develop a mode of
operation that will allow the input model to evolve to a specified
temperature (or luminosity) automatically. A small grid of starter models
with different masses covering the range of interest can be constructed
independently of the optimization. Other interesting parameters, like the
helium mass fraction ($Y$) and the metallicity ($Z$), can be specified at the
beginning of the evolution. When the model has evolved to the parameter
values requested by the genetic algorithm, the adiabatic pulsation
frequencies can be calculated and compared to the observed periods. This
will lead to a goodness-of-fit measure that the genetic algorithm will
attempt to maximize.

{\bf Optimizing the efficiency.} The efficiency of genetic-algorithm-based
optimization is defined as the number of model evaluations required to
yield the global solution, relative to the number of models that would be
required for enumerative search of the grid at the same sampling density.
In practice, a genetic algorithm is usually hundreds or even thousands of
times more efficient than a complete grid, and its performance is fairly
insensitive to the few internal parameters that control its operation. We
will initially set these internal parameters based on our experience with
white dwarf models, but we will run synthetic data through the
optimization procedure (a so-called `hare \& hound' exercise) to ensure
that the input parameters are recovered faithfully. We will repeat these
exercises with variations in the control parameters until the efficiency
of the algorithm is optimal.

\subsection{Application to $\delta$-Scuti Stars}

Extending the Cepheid and RR Lyrae instability strip down to the main
sequence, we find a group of A and early-F type stars with pulsation
periods between roughly half an hour and half a day. This class of
pulsating stars, with several dozen known members, has been given the name
of the prototype, $\delta$-Scuti. The amplitude of the observed variation
ranges from a few milli-magnitudes to a few tenths of a magnitude, and
they often exhibit a mixture of both radial and non-radial modes.

{\bf Parameter sampling density.} The interpretation of pulsation data for
$\delta$-Scuti stars is currently facing challenges quite similar to those
faced by white dwarf modelers five years ago. The observational
requirements for long-term photometric monitoring have been satisfied by
successful multi-site campaigns on several stars \citep[e.g.,
see][]{bre98,bre99,han00}. But exploration of the most important physical
parameters in theoretical models has been limited to very coarse grids,
making it difficult to establish a unique best-fit model for a particular
set of observations. The most extensive attempt at model-fitting to date,
in terms of the number of computed models, was published by \cite{pam98},
who explored three parameters with a grid of 120 models. Using this basic
grid, they attempted to interpolate the pulsation frequencies and model
parameters to produce a much finer grid with 40,000 points. In the end
they found that the initial grid density was inadequate, leading to
significant differences between the interpolated periods and those
resulting from a complete evolutionary calculation with the same
parameters.

We will use the genetic algorithm to explore $\delta$-Scuti models over a
broad range of masses $(M)$ and chemical compositions $(Y,Z)$, covering
the full extent of the instability strip and allowing roughly 100 possible
values for each parameter. With various assumptions about convective core
overshooting $(\alpha_{\rm ov})$ and rotation $(v_{\rm rot})$---or
including them as adjustable parameters if warranted---the genetic
algorithm method will effectively sample the parameters with 10-100 times
greater density than any previous model-fitting attempt. It will do so
without the necessity of calculating the complete grid, because it samples
primarily those areas of the model space that it objectively finds to
produce better fits to the observations.

{\bf Mode identification.} The problem with grid density has been
exacerbated by uncertainties in the identification of the spherical degree
and azimuthal order ($l,m$) of the pulsation modes. The pattern of hot
and cool regions on the surface of a non-radially pulsating star can be
decomposed into spherical harmonic functions described by these two
indices, leading to distinguishable patterns of radial velocity and light
variation. The frequency of the variation also depends on the radial
overtone ($n$), which is not directly observable. The excitation mechanism
for the pulsations observed in $\delta$-Scuti stars is not well
understood; only a small fraction of the pulsation modes that are
theoretically possible appear to be excited to detectable amplitudes. This
creates some difficulty for a unique interpretation of the observed
frequencies, a problem which has only recently started to be resolved
using multi-color photometric techniques \citep{dup03}. As space-based
data start to become available, allowing the reliable detection of more
low amplitude pulsation modes, this issue is likely to be less of a
problem. For example, recent WIRE observations of $\theta^2$ Tauri
\citep{por02} nearly tripled the number of pulsation modes detected in
this star, relative to earlier multi-site observations from the ground.

We will investigate the feasibility of incorporating mode identification
directly into the model-fitting procedure, whenever it has not been
determined independently from observations. The idea is to make no {\it a
priori} assumptions about the ($n,l,m$) values in the absence of
observational constraints, and just try to match the frequencies. Within
the range of frequencies where pulsations are observed, the number of
modes that are theoretically possible increases quickly with the value of
$l$; rotation and magnetic fields split non-radial modes into 2$l$+1
components. So the goodness-of-fit measure must be weighted to correct for
the mode density, greatly enhancing the fitness of models that match one
of the observed frequencies with a radial mode (because there are few) and
only slightly enhancing the fitness when an $l$=3 mode matches the
observations (because there are many).

{\bf Convective core overshooting.} Despite these difficulties,
$\delta$-Scuti stars are currently the most promising candidates for
asteroseismology near the main-sequence, and objective global
model-fitting to their pulsation frequencies promises to yield important
insights into their interior structure and evolutionary history. For
example, because some of these stars are slightly evolved, an exciting
possibility is to use them as a direct test for the presence and degree of
convective core overshooting, which is expected to leave a dramatic
imprint on the pulsation spectrum. \cite{tem01} showed that changes to the
convective core overshooting parameter in their models had a fundamentally
different effect on the theoretical pulsation frequencies than any of the
other parameters ($M, Y, Z$).

We will initially perform the model-fitting by considering only two
possible values for the core overshooting parameter: $\alpha_{\rm
ov}$=\,0.0 (no overshooting), and $\alpha_{\rm ov}$=\,0.2, the value
inferred by \cite{rib00} from an analysis of eclipsing binary data. This
will reveal the degree to which the frequencies can be fit better by a
model that includes convective overshooting, and it may motivate
additional fits that allow overshooting to be a fully adjustable
parameter. This is analogous to the way we approached the sensitivity of
white dwarf models to core composition: first allowing several discrete
values to demonstrate the potential of the parameter to improve the fits
\citep{mnw00}, and then following up with a full scale exploration leading
to an estimate that was much more precise than was earlier thought 
possible \citep{mwc01}.

\subsection{Solar-like Oscillations}

Solar-like oscillations have now been detected in two main-sequence stars
($\alpha$ Cen A and B), several sub-giants ($\eta$ Boo, Procyon, $\beta$
Hyi), and several giants ($\xi$ Hya, Arcturus, $\alpha$ UMa). The
oscillation amplitudes and the frequency of maximum power in these stars
agree reasonably well with our theoretical expectations. The field is
progressing very rapidly, and there is good reason to believe that many
new observations will become available in the next few years, particularly
after the launch of COROT and Kepler. With this in mind, it will be a
distinct advantage to have in place the analysis methods that can make
sense of these data efficiently, leading us quickly to a deeper
understanding of the solar oscillations in the context of similar
pulsations in other stars.

{\bf Extending the method.} After adapting the genetic algorithm fitting
method to main-sequence models for the analysis of $\delta$-Scuti stars,
it will be straightforward to extend the method to solar-like pulsators.
This primarily requires a redefinition of the range for each of the
relevant parameters. We have experience making this kind of transition
with white dwarf models---extending the genetic algorithm method from the
helium-atmosphere (DB) to the hydrogen-atmosphere (DA) white dwarfs by
simply changing the allowed temperature range and adding one adjustable
parameter for the hydrogen layer mass. Though not trivial, the time-scale
for this development was short compared to the time that it took to adapt
the model to interface with the parallel genetic algorithm.

A simplifying circumstance for the analysis of solar-like oscillations,
compared to $\delta$-Scuti pulsations, is the relative ease of mode
identification. The excitation mechanism for solar-like oscillations is
convection near the surface, creating a broad envelope of power with a
peak that scales with the acoustic cutoff frequency \citep{bro91}. Within
this envelope a large fraction of the theoretically possible pulsation
modes are excited to detectable amplitudes, leading to readily
identifiable patterns. Without any detailed modeling, these overall
patterns (the so-called large and small separations, $\Delta\nu$ and
$\delta\nu$) immediately lead to an estimate of the mean density of the
star and can indicate the presence of interior chemical gradients. But a
full analysis must include a detailed comparison of the individual
frequencies with theoretical models. With a sufficiently long time
baseline, it should also be possible to resolve the rotationally split
$m$-components of the non-radial ($l\ge1$) modes. The frequency
separation between these components can yield information about the
internal rotation rate, since each mode samples the interior in a slightly
different manner.

{\bf Hare \& Hound exercises.} Before any new observations are available,
we will perform theoretical investigations using synthetic data to
document the results expected to emerge from the various space missions,
and to benchmark and fine tune the optimization method with `hare \&
hound' exercises to maximize its efficiency. When new observations become
available that may benefit from the global exploration of models made
possible by the genetic algorithm, we will use this powerful tool to
extract the complete physical insight that the data can provide.

\section{Call for Collaborators}

The proposal outlined above will initially be supported for three years by
an NSF Fellowship at the High Altitude Observatory, to begin in the fall
of 2004. Comments, suggestions, and collaborators are welcome. Please
contact Travis Metcalfe $<$travis@hao.ucar.edu$>$ for more information
about the current status of the project.

\end{document}